\documentclass[prc,twocolumn,showpacs,amsmath,amssymb,superscriptaddress,floatfix,nofootinbib,10pt]{revtex4}

\usepackage[utf8]{inputenc}
\usepackage{mathrsfs,bm}
\usepackage{longtable,lscape}
\usepackage{txfonts}
\usepackage{amssymb}
\usepackage{indentfirst}
\usepackage{graphicx,booktabs}
\usepackage{multirow}
\usepackage{overpic}
\usepackage{color}
\usepackage{amssymb}
\usepackage{hyperref}
\usepackage{ulem}

\begin{document}
\title{Model independent determination of the spins of the  $P_{c}$(4440) and $P_{c}$(4457)  from the spectroscopy of the triply charmed dibaryons
}

\author{Ya-Wen Pan}
\affiliation{School of Physics and
Nuclear Energy Engineering \& International Research Center for Nuclei and Particles in the Cosmos \&
Beijing Key Laboratory of Advanced Nuclear Materials and Physics,  Beihang University, Beijing 100191, China}

\author{Ming-Zhu Liu}
\affiliation{School of Physics and
Nuclear Energy Engineering \& International Research Center for Nuclei and Particles in the Cosmos \&
Beijing Key Laboratory of Advanced Nuclear Materials and Physics,  Beihang University, Beijing 100191, China}

\author{Fang-Zheng Peng}
\affiliation{School of Physics and
Nuclear Energy Engineering \& International Research Center for Nuclei and Particles in the Cosmos \&
Beijing Key Laboratory of Advanced Nuclear Materials and Physics,  Beihang University, Beijing 100191, China}

\author{Mario {S\'anchez S\'anchez}}
\affiliation{Centre d'\'Etudes Nucl\'eaires, CNRS/IN2P3, Universit\'e de Bordeaux, 33175 Gradignan, France}

\author{Li-Sheng Geng}
\email[E-mail: ]{lisheng.geng@buaa.edu.cn} \affiliation{School of
Physics and Nuclear Energy Engineering \& International Research
Center for Nuclei and Particles in the Cosmos \& Beijing Key
Laboratory of Advanced Nuclear Materials and Physics,  Beihang
University, Beijing 100191, China}
\affiliation{School of Physics and Engineering, Zhengzhou University, Zhengzhou, Henan 450001, China}

\author{Manuel Pavon Valderrama}\email[E-mail: ]{mpavon@buaa.edu.cn}
\affiliation{School of Physics and Nuclear Energy Engineering \&
International Research Center for Nuclei and Particles in the
Cosmos \& Beijing Key Laboratory of Advanced Nuclear Materials and
Physics,  Beihang University, Beijing 100191, China}

\begin{abstract}

  The LHCb collaboration has recently observed three narrow pentaquark
  states --- the $P_c(4312)$, $P_c(4440)$, and $P_c(4457)$ ---
  that are located close to the $\bar{D} \Sigma_c$ and $\bar{D}^* \Sigma_c$
  thresholds.
  Among the so-far proposed theoretical interpretations for these pentaquarks,
  the molecular hypothesis seems to be the preferred one.
  Nevertheless, in the molecular picture the spins of the $P_c(4440)$
  and $P_c(4457)$ have not been unambiguously determined yet.
  In this letter we point out that heavy antiquark-diquark symmetry
  induces a model-independent relation between the spin-splitting
  in the masses of the $P_c(4440)$ and $P_c(4457)$ $\bar{D}^* \Sigma_c$
  pentaquarks and the corresponding splitting for the $0^+$ and $1^+$
  $\Xi_{cc} \Sigma_c$ triply charmed dibaryons.
  This is particularly relevant owing to a recent lattice QCD prediction of
  the $1^+$ triply charmed dibaryon, which suggests that a calculation of
  the mass of its $0^+$ partner might be within reach.
  This in turn would reveal the spins of the $P_c(4440)$ and $P_c(4457)$
  pentaquarks, providing a highly nontrivial test of heavy-quark
  symmetry and the molecular nature of the pentaquarks.
  Furthermore, the molecular interpretation of the hidden-charm
  pentaquarks implies the existence of a total of ten triply charmed
  dibaryons as $\Xi_{cc}^{(*)} \Sigma_c^{(*)}$ molecules,
  which, if confirmed in the lattice, will largely expand our understanding
  of the non-perturbative strong interaction in the heavy-quark sector.
\end{abstract}

\pacs{13.60.Le, 12.39.Mk,13.25.Jx}

\maketitle

Recently the LHCb collaboration has reported the observation of
three pentaquark states~\cite{Aaij:2019vzc} --- the $P_{c}(4312)$,
$P_{c}(4440)$ and $P_{c}(4457)$ --- very close to two meson-baryon
thresholds, the $\bar{D}\Sigma_{c}$ in the case of the $P_{c}(4312)$
and the $\bar{D}^{\ast}\Sigma_{c}$ in the case of the 
$P_{c}(4440)$ and $P_{c}(4457)$.
This interesting coincidence has been promptly interpreted
by a series of theoretical
works~\cite{Chen:2019bip,Chen:2019asm,Liu:2019tjn,Guo:2019fdo,Xiao:2019aya,Shimizu:2019ptd,Guo:2019kdc}
as evidence for their molecular nature, a hypothesis which is
further fueled by the predictions a few years ago of
the existence of these pentaquarks~\cite{Wu:2010jy,Wu:2010vk,Xiao:2013yca,Karliner:2015ina,Wang:2011rga,Yang:2011wz}.
In addition to  the molecular picture,
there are other explanations, e.g., hadro-charmonium~\cite{Eides:2019tgv}
or compact pentaquark states~\cite{Ali:2019npk,Wang:2019got,Cheng:2019obk}.
It is also worth noticing that the $P_c(4440)$ and $P_c(4457)$ were previously
identified as a single state, the $P_c(4450)$~\cite{Aaij:2015tga}.

However the molecular interpretation cannot unambiguously determine
the spins of the $P_c(4440)$ and $P_c(4457)$,
which can be either $1/2$ or $3/2$.
Initial studies showed a preference for the $P_c(4440)$ and $P_c(4457)$
being identified with $J=1/2$ and $3/2$ molecules
respectively~\cite{Chen:2019asm,Liu:2019tjn,Xiao:2019aya},
while recent theoretical works prefer the opposite
identification~\cite{Yamaguchi:2019seo,Valderrama:2019chc,Liu:2019zvb}.

Given this situation~\footnote{Even experiments might need
  larger statistics to give a conclusive answer.},
{one might turn to  lattice QCD for a model independent answer}.
But meson-baryon systems % with four quarks and one antiquark
are difficult to study directly in the lattice,
more so if there are coupled channels.
After the 2015 % LHCb discovery of the previous
pentaquark peaks~\cite{Aaij:2015tga},
there have been only two lattice QCD studies on $J/\psi N$ and $\eta_c N$
interactions~\cite{Sugiura:2019pye,Skerbis:2018lew},
none of which found pentaquarks.
These studies, which were performed within a single-channel approximation,
concluded that further calculations including coupled-channels were needed.

To circumvent the aforementioned difficulties, 
% Facing such a situation
we propose % in this work
to learn more about the $\bar{D}^{(*)} \Sigma_c^{(*)}$ pentaquarks from
the $\Xi_{cc}^{(*)} \Sigma_c^{(*)}$ dibaryons, {which
  are connected by heavy antiquark-diquark symmetry (HADS)}.

Hadronic molecules containing heavy quarks ($c$, $b$)
  are constrained by heavy-quark symmetry
  (HQS)~\cite{Isgur:1989vq,Isgur:1989ed},
which can be exploited to predict unobserved
states~\cite{Nieves:2012tt,Guo:2013xga,Xiao:2019gjd,Peng:2019wys}
or explain relations among experimentally known
states~\cite{Bondar:2011ev,Mehen:2011yh,Guo:2013sya}.
Two manifestations of HQS are relevant for the present work:
heavy-quark spin symmetry (HQSS) (the interaction
between heavy hadrons is independent of the orientation  of
the heavy quark spin) and HADS~\cite{Savage:1990di}
(a pair of heavy-quarks behaves like a heavy antiquark).
Regarding the pentaquark states, HQSS predicts the existence of seven molecular states~\cite{Xiao:2013yca,Liu:2019tjn,Xiao:2019aya,Sakai:2019qph,Yamaguchi:2019seo,Valderrama:2019chc}.
Namely, from HQSS and the LHCb pentaquarks~\cite{Aaij:2019vzc}
one expects four more unobserved hidden-charm pentaquarks.

But the latest discovery of the LHCb pentaquarks leads to more consequences.
  From HADS, we expect the light-quark structures of the ${\bar D}^{(*)}$ and the ${\Xi}_{cc}^{(*)}$  to be almost
  identical: if the $\bar{D}^{(*)} \Sigma_c^{(*)}$ system binds,
  the ${\Xi}_{cc}^{(*)} \Sigma_c^{(*)}$ system should also bind.
  Specifically we can update the triply charmed dibaryon spectrum,
from the four dibaryons originally predicted in Ref.~\cite{Liu:2018zzu}
to ten dibaryons in the present manuscript.
As we will show,
this could help determine the spins of the $P_c(4440)$ and $P_c(4457)$.
Besides,
it seems to be much easier for lattice QCD
to study the $\Xi_{cc}^{(*)} \Sigma_c^{(*)}$ systems
than the $\bar{D}^{(*)}\Sigma_c^{(*)}$  systems,
which up to now lattice QCD studies have
not been able to simulate.
On the other hand, a recent lattice QCD study~\cite{Junnarkar:2019equ}
has reported the likely existence of a triply charmed $\Xi_{cc} \Sigma_c$
dibaryon with spin-parity $J^P = 1^+$,  isospin $I=\tfrac{1}{2}$ and
a binding energy of $ B_2 = (8 \pm 17) \,{\rm MeV} $,
where Ref.~\cite{Junnarkar:2019equ} indicates
that owing to the large uncertainties its existence is not guaranteed.
Here we predict this dibaryon to have a binding energy
of $ B_2 \sim (15-30)\,{\rm MeV}$,
which is compatible within its large error with % the lattice QCD study
the results of Ref.~\cite{Junnarkar:2019equ}, but clearly leans towards
the conclusion that the $1^+$ $\Xi_{cc} \Sigma_c$ dibaryon binds.
More importantly, as we will show below,
if lattice QCD is able to calculate
the mass of the $J^P = 0^+$ configuration or the mass splitting between the
$0^+$ and $1^+$ $\Xi_{cc}\Sigma_c$ dibaryons, it will provide a model-independent
determination of the spins of the $P_c(4440)$ and $P_c(4457)$. 

Now we  explain how we predict the binding energies
of the triply charmed dibaryons.
For this we construct a contact-range effective field theory (EFT)
that respects HQS, in the line of Refs.~\cite{Liu:2018bkx,Liu:2019tjn}.
We will follow a streamlined approach, where the starting point is
the ``brown muck'' within the heavy hadrons,
i.e., the light-quark subfields.
The idea is that the heavy quark is merely a spectator, therefore it is more economic
to write the interaction between the heavy hadrons in terms of
the light-quark degrees of freedom~\cite{Valderrama:2019sid}.
Thus, a heavy antimeson can be written as a non-relativistic light-quark
subfield $q_L$.
Owing to HADS the ``brown muck'' in a doubly heavy baryon is basically
the same as that in a heavy antimeson, i.e., we can represent the doubly
heavy baryon with the same light-quark subfield $q_L$
as the heavy antimeson.
Finally, the (singly) heavy baryons can be represented with a $S=1$
light-diquark subfield $d_L$.

With these pieces, we can write the lowest-dimensional, contact-range
Lagrangian involving a light-quark and light-diquark subfield as
\begin{eqnarray}
  \mathcal{L}_{\rm contact} &=& 
  C_a \, (q_L^{\dagger}\,q_L) \, (d_L^{\dagger}\,d_L)
  \nonumber \\
  &+& C_b\,(q_L^{\dagger}\,\vec{\sigma}_{L}\,q_L) \, \cdot \,
  (d_L^{\dagger}\,\vec{S}_{L}\,d_L) \, , \label{eq:lagrangian-L}
\end{eqnarray}
where $\vec{\sigma}_L$ are the Pauli matrices for the $q_L$ subfield and
$\vec{S}_L$ the spin-1 matrices for the $d_L$ subfield.
From this Lagrangian, we end up with the non-relativistic potential
\begin{eqnarray}
  V_C = C_a + C_b \, \vec{\sigma}_L \cdot \vec{S}_L \, . \label{eq:pot-L}
\end{eqnarray}
This is supplemented by a series of rules to translate
the light-quark spin operators into the heavy hadron ones.
For the heavy antimeson fields we have
\begin{eqnarray}
  \langle {\bar D} | \vec{\sigma}_L | {\bar D} \rangle &=& 0 % \, , \\
  \quad , \quad
  \langle {\bar D}^* | \vec{\sigma}_L | {\bar D}^* \rangle = \vec{S}_1 \, ,
\end{eqnarray}
 where $\vec{S}_1$ are the spin-1 matrices as applied to
 the ${\bar D}^*$ meson.
For the doubly heavy baryons
\begin{eqnarray}
  \langle \Xi_{cc} | \vec{\sigma}_L | \Xi_{cc} \rangle &=&
  -\frac{1}{3}\,\vec{\sigma}_1 % \, ,  \label{eq:Xicc-L} \\
  \quad , \quad
  \langle \Xi_{cc}^* | \vec{\sigma}_L | \Xi_{cc}^* \rangle =
  +\frac{2}{3}\,\vec{\Sigma}_1 \, , \label{eq:Xicc-L} 
\end{eqnarray}
where $\vec{\sigma}_1$ and $\vec{\Sigma}_1$ are the Pauli and
spin-$\tfrac{3}{2}$ matrices as applied to
the $\Xi_{cc}$ and $\Xi_{cc}^*$ baryons.
Finally for the singly heavy hadrons
% containing a spin-1, S-wave light-diquark we have
\begin{eqnarray}
  \langle \Sigma_{c} | \vec{S}_L | \Sigma_{c} \rangle &=&
  +\frac{2}{3}\,\vec{\sigma}_2 % \, , \\
  \quad , \quad
  \langle \Sigma_{c}^* | \vec{S}_L | \Sigma_{c}^* \rangle =
  +\frac{2}{3}\,\vec{\Sigma}_2 \, ,
\end{eqnarray}
with $\vec{\sigma}_2$ and $\vec{\Sigma}_2$ as in Eq.~(\ref{eq:Xicc-L}),
but applied to the $\Sigma_{c}$ and
$\Sigma_{c}^*$ baryons.
With these rules and the contact-range potentials of Eq.~(\ref{eq:pot-L})
we arrive at the potentials of Table~\ref{tab:potential}.
We only consider contact-range interactions, as
previous studies of heavy hadron-hadron
systems~\cite{Valderrama:2012jv,Lu:2017dvm},
%and heavy baryon-baryon systems
indicate that pion exchanges are perturbative
%for the heavy antimeson-baryon systems
in the charm sector.

The potentials of Table~\ref{tab:potential} are subject to a certain
degree of uncertainty. They are derived from HQS, which is only exact
in the limit of infinite heavy-quark mass, $m_Q \to \infty$.
For finite $m_Q$ we expect small violations, which for HQSS are of the order
of $\Lambda_{\rm QCD} / m_Q$ with $\Lambda_{QCD} \sim (200-300)\,{\rm MeV}$
the QCD scale, i.e., we expect a $15\%$ error in the charm sector
for the pentaquark potentials.
For HADS, the uncertainty is of the order of $\Lambda_{QCD} / (m_Q v)$~\cite{Savage:1990di}, with $v$ the expected velocity for the heavy-quark pair.
From the estimation of Ref.~\cite{Hu:2005gf}, $m_Q v \sim 0.8\,{\rm GeV}$
for a charm quark pair, we arrive at a $25-40\%$ uncertainty for HADS.
This figure has been conjectured to be larger,
up to the point that HADS might not be applicable
in the charm sector~\cite{Cohen:2006jg}.
But recent lattice QCD calculations~\cite{Padmanath:2015jea,Chen:2017kxr,Alexandrou:2017xwd,Mathur:2018rwu} of the mass splitting between
the $J=\frac{1}{2}$ and $J=\frac{3}{2}$
doubly charmed baryons are usually $20-25\%$ smaller
than the HADS prediction,
%(a splitting in the $70-80\,{\rm MeV}$ range,
%to be compared with $3 (m(D^*)-m(D))/4 \sim 105\,{\rm MeV}$~\cite{Hu:2005gf}),
suggesting % that HADS works considerably well.
our lower estimation for the HADS uncertainty.
From this we will settle on a $25\%$ error
for the dibaryon potentials.

\begin{table}[!h]
  \centering \caption{The lowest-order contact range potentials
    for the heavy antimeson - heavy baryon and doubly heavy
    baryon - heavy baryon systems,
    which depend on two unknown coupling constants $C_{a}$ and $C_{b}$.
\label{tab:potential}}
\begin{tabular}{ccc|cccccccc}
  \hline\hline state& $J^{P}$ &V   & state   &$J^{P}$   &V\\
  
  \hline \multirow{2}{0.8cm}{$\bar{D}\Sigma_{c}$} &
  \multirow{2}{0.8cm}{$1/2^{-}$} &\multirow{2}{0.8cm}{$C_{a}$}
  & \multirow{2}{0.8cm}{$\Xi_{cc}\Sigma_{c}$}
  & $0^{+}$  & $C_{a}+\frac{2}{3}C_{b}$          \\
  & & & & $1^{+}$    & $C_{a}-\frac{2}{9}C_{b}$           \\
  \hline
  \multirow{2}{0.8cm}{$\bar{D}\Sigma_{c}^{\ast}$}&
  \multirow{2}{0.8cm}{$3/2^{-}$} &\multirow{2}{0.8cm}{$C_{a}$}
  &\multirow{2}{0.8cm}{$\Xi_{cc}\Sigma_{c}^{\ast}$}
  & $1^{+}$  & $C_{a}+\frac{5}{9}C_{b}$          \\
  & & & & $2^{+}$    & $C_{a}-\frac{1}{3}C_{b}$           \\
  \hline
  \multirow{2}{0.8cm}{$\bar{D}^{\ast}\Sigma_{c}$}&$1/2^{-}$
  &$C_{a}-\frac{4}{3}C_{b}$ &
  \multirow{2}{0.8cm}{$\Xi_{cc}^{\ast}\Sigma_{c}$}  & $1^{+}$  & $C_{a}-\frac{10}{9}C_{b}$
\\
& $3/2^{-}$ &$C_{a}+\frac{2}{3}C_{b}$
 & & $2^{+}$  & $C_{a}+\frac{2}{3}C_{b}$
\\
\hline
\multirow{4}{0.8cm}{$\bar{D}^{\ast}\Sigma_{c}^{\ast}$} & $1/2^{-}$
&$C_{a}-\frac{5}{3}C_{b}$  &
\multirow{4}{0.8cm}{$\Xi_{cc}^{\ast}\Sigma_{c}^{\ast}$}  &$0^{+}$
& $C_{a}-\frac{5}{3}C_{b}$
\\
& \multirow{2}{0.8cm}{$3/2^{-}$}
& \multirow{2}{1.5cm}{$C_{a}-\frac{2}{3}C_{b}$} &
& $1^{+}$ & $C_{a}-\frac{11}{9}C_{b}$
\\  & &  &
& $2^{+}$ & $C_{a}-\frac{1}{3}C_{b}$
\\
& {$5/2^{-}$}    &{$C_{a}+C_{b}$}    &   & $3^{+}$ & $C_{a}+C_{b}$    \\
\hline\hline
\end{tabular}
\end{table}

To obtain concrete predictions %for molecular states
we have to solve a non-relativistic bound state equation
with the contact-range potentials of Table \ref{tab:potential}.
If we work in momentum space we simply solve the Lippmann-Schwinger equation
for the bound state pole, which reads
\begin{eqnarray}
\phi(k)+\int \frac{d^{3}p}{(2\pi)^3}\langle
k|V|p\rangle\frac{\phi(p)}{B+\frac{p^2}{2\mu}}=0,
\label{10}
\end{eqnarray}
where $\phi(k)$ is the vertex function, $B$ the binding energy and $\mu$
the reduced mass.
To solve this equation % this bound state equation,
% first we have to regularize the contact potential
we first regularize the potential
\begin{eqnarray}
\langle p| V_{\Lambda}|
p^{\prime}\rangle=C(\Lambda)f(\frac{p}{\Lambda})f(\frac{p^{\prime}}{\Lambda}),
\end{eqnarray}
with $\Lambda$ the cutoff, $f(x)$ a regulator, % function,
and $C$ the linear combination of the $C_a$ and $C_b$ couplings % constants
that corresponds to the particular molecular state we are interested in.
Notice that $C = C(\Lambda)$, i.e., the couplings depend on the cutoff,
which is necessary if we want to properly renormalize the calculations,
i.e., we want the predictions to depend on  the cutoff only moderately,
where cutoff variations represent the uncertainty coming
from subleading order terms that we have not explicitly
taken into account.
We will choose a Gaussian regulator of the type $f(x) = e^{-x^2}$ and
for the cutoff we will use the range $\Lambda=0.5-1$ GeV,
where the average within this range % can be identified with
corresponds to the $\rho$ meson mass.

For the masses of heavy hadrons, we use
$m_{D}=1867$ MeV, $m_{D^{\ast}}=2009$ MeV, $m_{\Sigma_{c}}=2454$ MeV,
$m_{\Sigma_{c}^{\ast}}=2518$ MeV,
$m_{\Xi_{cc}}=3621$ MeV, and $m_{\Xi_{cc}^{\ast}}=3727$ MeV, where
the mass of $\Xi_{cc}^{\ast}$ has been deduced from the HADS relation
$m_{\Xi_{cc}^{\ast}}-m_{\Xi_{cc}}= 3\,(m_{D^{\ast}}-m_{D})/4$~\cite{Hu:2005gf}.\footnote{
Ref.~\cite{Mehen:2019cxn} has calculated perturbative corrections
to the above mass relation, which happen to be small ($\approx10\%$) and 
we will ignore here.}

\begin{figure*}[htpb]
\centering
\includegraphics[scale=.5]{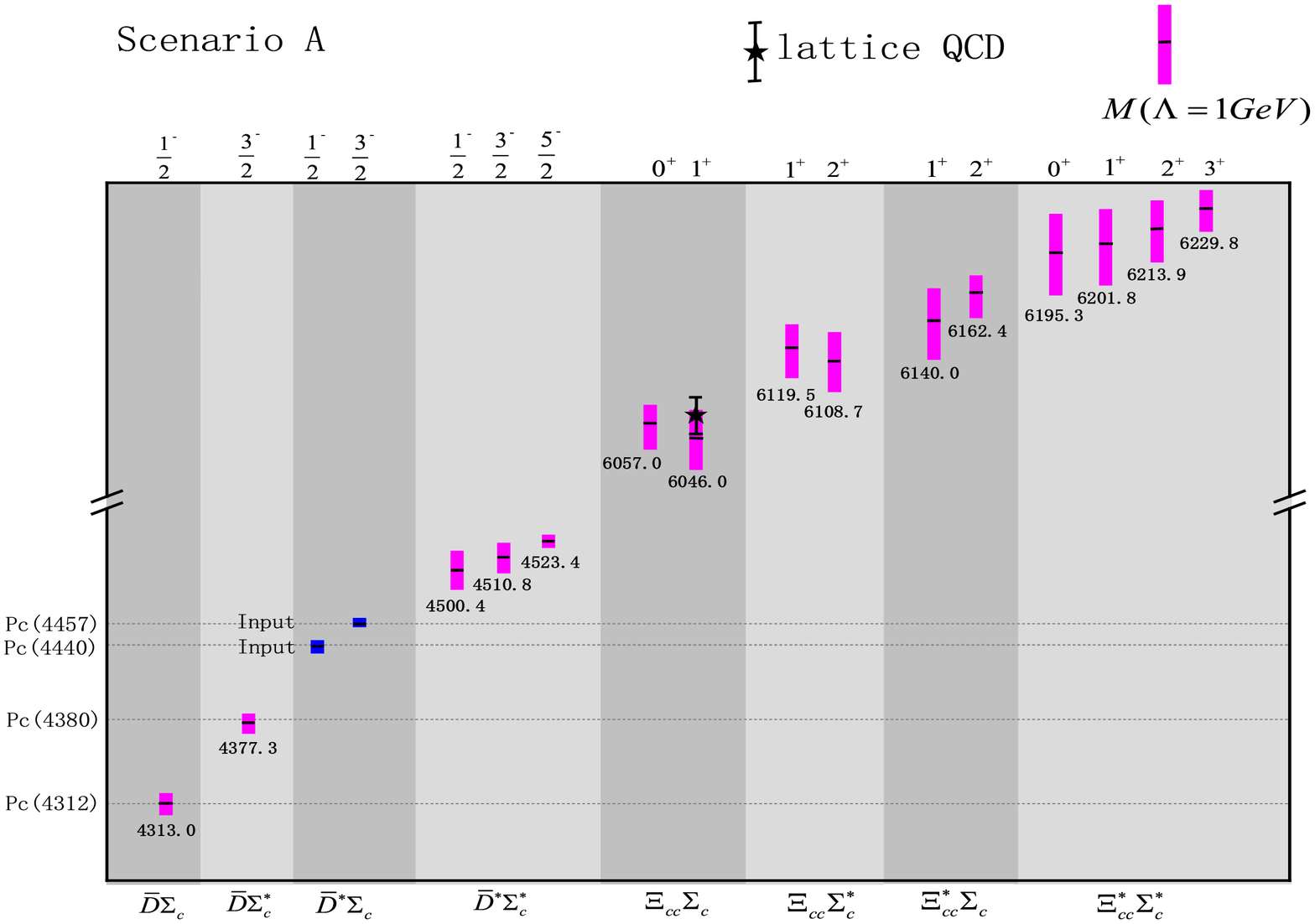}
\includegraphics[scale=.5]{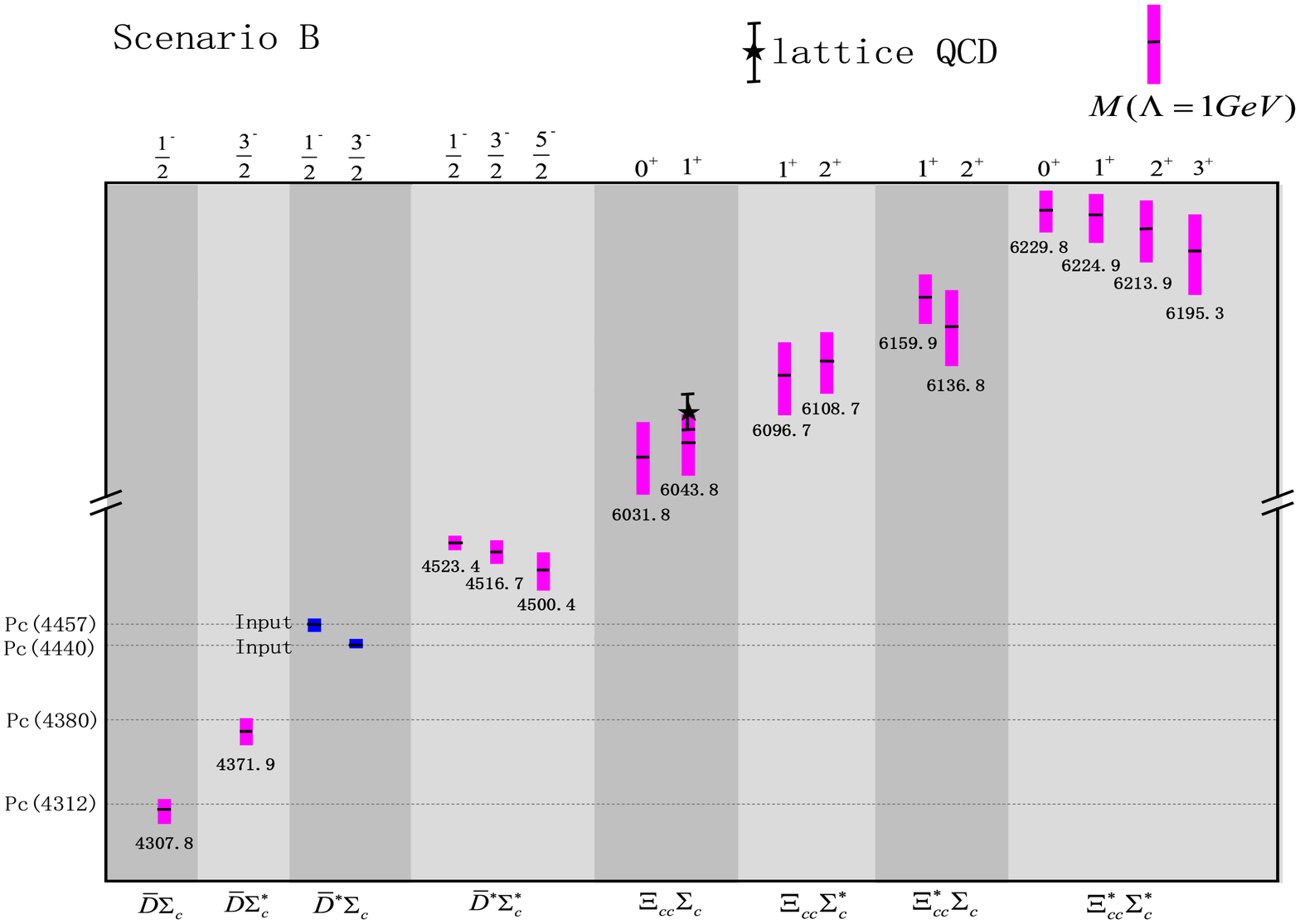}
\caption{Predicted masses   of pentaquark states and hexaquark (dibaryon) states
 in scenario A (upper), in which the $P_c(4440)$/$P_c(4457)$
  is identified with the $J=\tfrac{1}{2}$/$\tfrac{3}{2}$ $\bar{D}^* \Sigma_c$
  molecule, and scenario B (lower), where the opposite assignment is made.
  The dotted horizontal lines indicate the central experimental masses of
  the $P_c(4312)$,  $P_c(4380)$, $P_c(4440)$, and $P_c(4457)$. The black solid lines denote the masses obtained with $\Lambda=1$ GeV
  while the red bands denote the uncertainties originated from HQS ($15\%$) and HADS ($25\%)$ breaking. The lattice QCD result for the $1^+$ $\Xi_{cc}\Sigma_c$ state is
  calculated using the binding energy of Ref.~\cite{Junnarkar:2019equ} and the experimental masses for $\Xi_{cc}$ and $\Sigma_c$.}
\label{fig:spe}
\end{figure*}

Now we still have to determine the couplings $C_a$ and $C_b$.
For this, we notice that if the $P_c(4440)$ and $P_c(4457)$ are indeed
$\bar{D}^* \Sigma_c$ bound states, their contact-range potentials read
\begin{eqnarray}
  V(\bar{D}^* \Sigma_c, J = \tfrac{1}{2}) &=& C_a - \frac{4}{3}\,C_b \, , \\
  V(\bar{D}^* \Sigma_c, J = \tfrac{3}{2}) &=& C_a + \frac{2}{3}\,C_b \, .
\end{eqnarray}
Since the spins of the hidden-charm pentaquarks
are not experimentally known yet,
there are two possible identifications: the $P_c(4440)$ is
the spin $\tfrac{1}{2}$ state and the $P_c(4457)$
the spin $\tfrac{3}{2}$ one (scenario A),
or vice versa (scenario B)~\cite{Liu:2019tjn}.
This in principle give us a consistency test in terms of the postdiction
of the $P_c(4312)$ as a $\bar{D} \Sigma_c$ bound state,
see Table \ref{tab:predictions-Pc} for  details.
The outcome is inconclusive though: scenario A is marginally preferred
over scenario B.
In both scenarios,
we arrive at the conclusion that the ten possible triply
charmed dibaryons bind, with binding energies
of $10-50\,{\rm MeV}$, 
  see Fig.~\ref{fig:spe} 
for a graphical representation of the full dibaryon
spectrum and more details can be found in Table \ref{tab:predictions}.

One immediately notices that there is a strong correlation between
the ordering of the triply charmed baryons and that of
the pentaquark states, particularly
that of $P_c(4440)$ and $P_c(4457)$.
For instance, if the $0^+$ $\Sigma_c\Xi_{cc}$ state had a larger mass
than its $1^+$ counterpart, then scenario A would be preferred.
The difference in their binding energies is about 10 MeV,
which %may not be too difficult to achieve on the lattice.
might be achievable in the lattice.
In addition, one can study (one of) the three extra multiplets to
confirm the conclusion.
For instance, a decrease of the mass as a function of the spin
in the $\Xi_{cc}\Sigma_c$ $(0^+,1^+)$ multiplet and an increase
in the $\Xi_{cc}^*\Sigma_c$ $(1^+,2^+)$ multiplet
will both be unambiguous signals
that scenario A is preferred.

More concretely, from Table I and assuming that the $C_b$ coupling
is perturbative,
we arrive at the relation
\begin{eqnarray}
  M(1^+)-M(0^+) &=& - \frac{4}{9} \left( M(\tfrac{3}{2}^-)-M(\tfrac{1}{2}^-)
  \right)
  \nonumber \\
  % + \mathcal{O}(C_b^2) \nonumber \\
  &\approx& \mp \, (7.6 \pm 1.9)\,\mathrm{MeV} \, , \label{eq:mass-splitting}
\end{eqnarray}
with the $-/+$ sign respectively for scenarios A and B
and where we have used the experimental pentaquark masses to obtain
the number on the second line (with the error corresponding to
the HADS uncertainty).
It is worth noticing that in our numerical study
with % a cutoff of 0.5 GeV, we obtain  $M(1^+)-M(0^+)=-8.4$ MeV
$\Lambda = 0.5\,{\rm GeV}$, we obtain  $M(1^+)-M(0^+)=-8.4$ MeV
and $+9.7$ MeV for scenarios A and B, respectively.

In summary, the experimental discovery of the three pentaquark states
--- the $P_c(4312)$, $P_c(4440)$ and $P_c(4457)$ ---
which are conjectured to be hadronic molecules,
not only makes it possible to predict the full spectrum of
S-wave triply charmed dibaryons ($\Xi_{cc}^{(*)} \Sigma_c^{(*)}$ molecules)
but also opens the possibility of a model-independent determination of
the spins of the pentaquarks from lattice QCD simulations
of the dibaryons, which are definitively feasible
in the near future.
The basis for the prediction of the triply charmed dibaryons
are HQSS and HADS.
We exploited these two symmetries within the framework of a contact-range EFT,
which offers us a series of theoretical advantages including
systematicity and controlled error estimations.
The binding energies of the triply charmed dibaryons are predicted to
lie in the $(10-50)\,{\rm MeV}$ range, where the details
depend on the cutoff and the spins of the pentaquarks.
Among them, we predict a $1^+$ $\Xi_{cc} \Sigma_c$ dibaryon
with a binding energy of $15-30\,{\rm MeV}$, to be compared with 
$(8 \pm 17)\,{\rm MeV}$ in the lattice~\cite{Junnarkar:2019equ}.
The spin-splitting of the dibaryon masses together with HADS,
i.e., Eq.~(\ref{eq:mass-splitting}), indicates that a lattice
QCD calculation of the dibaryon spectrum will provide
a model-independent determination of the quantum numbers
of the LHCb pentaquarks.
The recent calculation of Ref.~\cite{Junnarkar:2019equ} implies that
such studies are within reach of current state of the art
lattice QCD simulations.

\begin{table*}[htpb]
  \caption{
    Binding energies for the pentaquarks composed of
    a charmed baryon and a charmed antimeson in the contact-range EFT
    defined in Eqs.~(\ref{eq:lagrangian-L}) and (\ref{eq:pot-L}).     %
    The two coupling constants, $C_a$ and $C_b$, are determined by
    reproducing the $P_c(4440)$ and $P_c(4457)$ as $\tfrac{1}{2}^-$
    and $\tfrac{3}{2}^-$ $\bar{D}^* \Sigma_c$ molecules (scenario A) or
    $\tfrac{3}{2}^-$ and $\tfrac{1}{2}^-$ $\bar{D}^* \Sigma_c$
    molecules (scenario B), respectively.
    The results are obtained with a cutoff of
    $\Lambda = 0.5(1.0)\,{\rm GeV}$.
    \label{tab:predictions-Pc}
}
\centering
\begin{tabular}{ccccccccccccc}
  \hline\hline State& $J^{P}$ &Threshold & $\Lambda$(GeV)
  & Scenario A  &  Scenario B    \\
\hline $P_{c}(4312)$($\bar{D}\Sigma_{c}$) & $\tfrac{1}{2}^{-}$ &4320.8&  0.5(1)
  &  $8.8^{+5.0}_{-4.2}$ $(7.6^{+9.2}_{-6.0})$ &  $14.4^{+6.5}_{-5.8}$ $(12.8^{+11.8}_{-8.6})$
  \\
  \hline $P_{c}(?)$($\bar{D}\Sigma_{c}^{\ast}$) &$\tfrac{3}{2}^{-}$ &4385.3& 0.5(1)
  &$9.1^{+5.0}_{-4.3}$ $(8.1^{+9.4}_{-6.3})$ & $14.7^{+6.6}_{-5.9}$ $(13.5^{+12.0}_{-8.9})$
  \\ \hline
  $P_{c}(4440)$($\bar{D}^{\ast}\Sigma_{c}$) &? &4462.2& 0.5(1)  &
  Input($\tfrac12^{-}$)     &Input($\tfrac32^{-}$)
  \\
  \hline $P_{c}(4457)$($\bar{D}^{\ast}\Sigma_{c}$)& ?
  &4462.2& 0.5(1) & Input($\tfrac32^{-}$) & Input($\tfrac12^{-}$)
  &
  \\
  \hline $P_{c}(?)$($\bar{D}^{\ast}\Sigma_{c}^{\ast}$) &$\tfrac12^{-}$ &4526.7& 0.5(1)
  & $25.6^{+9.0}_{-8.4}$ $(26.3^{+16.6}_{-13.7})$ & $3.0^{+2.8}_{-2.0}$ $(3.4^{+6.3}_{-3.2})$
  \\
  \hline $P_{c}(?)$($\bar{D}^{\ast}\Sigma_{c}^{\ast}$) &$\tfrac32^{-}$ &4526.7& 0.5(1) 
  & $15.8^{+6.7}_{-6.1}$ $(15.9^{+12.8}_{-9.8})$ & $10.0^{+5.2}_{-4.5}$ $(10.0^{+10.2}_{-7.2})$
  \\
  \hline $P_{c}(?)$($\bar{D}^{\ast}\Sigma_{c}^{\ast}$) &$\tfrac52^{-}$&4526.7& 0.5 (1) 
  & $3.0^{+2.8}_{-2.0}$ $(3.4^{+6.3}_{-3.2})$ & $25.6^{+9.0}_{-8.4}$ $(26.3^{+16.6}_{-13.7})$ 
\\ \hline \hline
\end{tabular}
\end{table*}
\begin{table*}[htpb]
  \caption{Same as Table \ref{tab:predictions-Pc} but for  the dibaryons composed of a doubly charmed
    baryon ($\Xi_{cc}$, $\Xi_{cc}^*$) and a singly charmed baryon
    ($\Sigma_c$, $\Sigma_c^*$).
    \label{tab:predictions}
  }
\centering
\begin{tabular}{ccccccccccccc}
\hline\hline State& $J^{P}$ &Threshold & $\Lambda$(GeV)   & Scenario(A)  &  Scenario(B)    \\
\hline $\Xi_{cc}\Sigma_{c}$ &$0^{+}$ &6074.9& 0.5(1)
& $10.0^{+7.9}_{-6.4}$ $(17.9^{+20.8}_{-14.4})$&  $30.4^{+15.5}_{-14.2}$ $(43.2^{+33.1}_{-27.2})$
\\
$\Xi_{cc}\Sigma_{c}$ &$1^{+}$ &6074.9&  0.5(1)
&  $18.4^{+11.3}_{-9.9}$ $(28.3^{+26.3}_{-20.1})$&  $20.7^{+12.1}_{-10.8}$ $(31.2^{+27.7}_{-21.6})$
\\ 
\hline $\Xi_{cc}\Sigma_{c}^{\ast}$ &$1^{+}$ &6139.5& 0.5(1)
&  $11.3^{+8.4}_{-7.0}$ $(20.0^{+25.4}_{-17.2})$&  $29.6^{+15.1}_{-13.8}$ $(42.8^{+32.7}_{-26.9})$
\\
  $\Xi_{cc}\Sigma_{c}^{\ast}$ &$2^{+}$&6139.5&0.5(1) & 
$20.0^{+11.8}_{-10.4}$ $(30.7^{+27.2}_{-21.2})$ &$20.0^{+11.8}_{-10.4}$ $(30.8^{+27.3}_{-21.3})$
\\
\hline   $\Xi_{cc}^{\ast}\Sigma_{c}$ &$1^{+}$ &6180.9& 0.5(1)&
$28.2^{+14.7}_{-13.4}$ $(41.0^{+32.0}_{-26.0})$ & $12.2^{+8.8}_{-7.4}$ $(21.0^{+22.4}_{-16.2})$
\\
 $\Xi_{cc}^{\ast}\Sigma_{c}$ &$2^{+}$ &6180.9& 0.5(1)&
$10.2^{+8.0}_{-6.5}$ $(18.5^{+21.0}_{-14.8})$ &$30.7^{+15.5}_{-14.2}$ $(44.1^{+33.3}_{-27.5})$
\\
\hline $\Xi_{cc}^{\ast}\Sigma_{c}^{\ast}$ &$0^{+}$ &6245.5& 0.5(1)&
$35.0^{+16.8}_{-15.6}$ $(50.2^{+35.6}_{-29.9})$ & $7.6^{+6.7}_{-5.3}$ $(15.8^{+19.2}_{-13.0})$
\\
 $\Xi_{cc}^{\ast}\Sigma_{c}^{\ast}$ &$1^{+}$ &6245.5& 0.5(1)&
$29.9^{+15.2}_{-13.9}$ $(43.7^{+32.9}_{-27.2})$ & $11.5^{+8.5}_{-7.0}$ $(20.7^{+22.0}_{-15.9})$
\\
 $\Xi_{cc}^{\ast}\Sigma_{c}^{\ast}$ &$2^{+}$ &6245.5&0.5(1)&
$20.3^{+11.8}_{-10.5}$ $(31.6^{+27.5}_{-21.6})$ & $20.3^{+11.8}_{-10.5}$ $(31.6^{+27.5}_{-21.6})$
\\
 $\Xi_{cc}^{\ast}\Sigma_{c}^{\ast}$ &$3^{+}$ &6245.5& 0.5(1)&
$7.6^{+6.7}_{-5.3}$ $(15.8^{+19.2}_{-13.0})$ & $35.0^{+16.8}_{-15.6}$ $(50.2^{+35.6}_{-30.0})$
\\
\hline\hline
\end{tabular}
\end{table*}

\section{Acknowledgments}
 This work is partly supported by the National Natural Science Foundation of China under Grant No.11735003, the fundamental Research Funds
for the Central Universities, and the Thousand
Talents Plan for Young Professionals.

\bibliography{triply-charmed-dibaryons}

\begin{thebibliography}{48}
\expandafter\ifx\csname natexlab\endcsname\relax\def\natexlab#1{#1}\fi
\expandafter\ifx\csname bibnamefont\endcsname\relax
  \def\bibnamefont#1{#1}\fi
\expandafter\ifx\csname bibfnamefont\endcsname\relax
  \def\bibfnamefont#1{#1}\fi
\expandafter\ifx\csname citenamefont\endcsname\relax
  \def\citenamefont#1{#1}\fi
\expandafter\ifx\csname url\endcsname\relax
  \def\url#1{\texttt{#1}}\fi
\expandafter\ifx\csname urlprefix\endcsname\relax\def\urlprefix{URL }\fi
\providecommand{\bibinfo}[2]{#2}
\providecommand{\eprint}[2][]{\url{#2}}

\bibitem[{\citenamefont{Aaij et~al.}(2019)}]{Aaij:2019vzc}
\bibinfo{author}{\bibfnamefont{R.}~\bibnamefont{Aaij}} \bibnamefont{et~al.}
  (\bibinfo{collaboration}{LHCb}), \bibinfo{journal}{Phys. Rev. Lett.}
  \textbf{\bibinfo{volume}{122}}, \bibinfo{pages}{222001}
  (\bibinfo{year}{2019}), \eprint{1904.03947}.

\bibitem[{\citenamefont{Chen et~al.}(2019{\natexlab{a}})\citenamefont{Chen,
  Chen, and Zhu}}]{Chen:2019bip}
\bibinfo{author}{\bibfnamefont{H.-X.} \bibnamefont{Chen}},
  \bibinfo{author}{\bibfnamefont{W.}~\bibnamefont{Chen}}, \bibnamefont{and}
  \bibinfo{author}{\bibfnamefont{S.-L.} \bibnamefont{Zhu}}
  (\bibinfo{year}{2019}{\natexlab{a}}), \eprint{1903.11001}.

\bibitem[{\citenamefont{Chen et~al.}(2019{\natexlab{b}})\citenamefont{Chen,
  Sun, Liu, and Zhu}}]{Chen:2019asm}
\bibinfo{author}{\bibfnamefont{R.}~\bibnamefont{Chen}},
  \bibinfo{author}{\bibfnamefont{Z.-F.} \bibnamefont{Sun}},
  \bibinfo{author}{\bibfnamefont{X.}~\bibnamefont{Liu}}, \bibnamefont{and}
  \bibinfo{author}{\bibfnamefont{S.-L.} \bibnamefont{Zhu}}
  (\bibinfo{year}{2019}{\natexlab{b}}), \eprint{1903.11013}.

\bibitem[{\citenamefont{Liu et~al.}(2019{\natexlab{a}})\citenamefont{Liu, Pan,
  Peng, Sánchez~Sánchez, Geng, Hosaka, and Pavon~Valderrama}}]{Liu:2019tjn}
\bibinfo{author}{\bibfnamefont{M.-Z.} \bibnamefont{Liu}},
  \bibinfo{author}{\bibfnamefont{Y.-W.} \bibnamefont{Pan}},
  \bibinfo{author}{\bibfnamefont{F.-Z.} \bibnamefont{Peng}},
  \bibinfo{author}{\bibfnamefont{M.}~\bibnamefont{Sánchez~Sánchez}},
  \bibinfo{author}{\bibfnamefont{L.-S.} \bibnamefont{Geng}},
  \bibinfo{author}{\bibfnamefont{A.}~\bibnamefont{Hosaka}}, \bibnamefont{and}
  \bibinfo{author}{\bibfnamefont{M.}~\bibnamefont{Pavon~Valderrama}},
  \bibinfo{journal}{Phys. Rev. Lett.} \textbf{\bibinfo{volume}{122}},
  \bibinfo{pages}{242001} (\bibinfo{year}{2019}{\natexlab{a}}),
  \eprint{1903.11560}.

\bibitem[{\citenamefont{Guo et~al.}(2019)\citenamefont{Guo, Jing, Meißner, and
  Sakai}}]{Guo:2019fdo}
\bibinfo{author}{\bibfnamefont{F.-K.} \bibnamefont{Guo}},
  \bibinfo{author}{\bibfnamefont{H.-J.} \bibnamefont{Jing}},
  \bibinfo{author}{\bibfnamefont{U.-G.} \bibnamefont{Meißner}},
  \bibnamefont{and} \bibinfo{author}{\bibfnamefont{S.}~\bibnamefont{Sakai}},
  \bibinfo{journal}{Phys. Rev.} \textbf{\bibinfo{volume}{D99}},
  \bibinfo{pages}{091501} (\bibinfo{year}{2019}), \eprint{1903.11503}.

\bibitem[{\citenamefont{Xiao et~al.}(2019{\natexlab{a}})\citenamefont{Xiao,
  Nieves, and Oset}}]{Xiao:2019aya}
\bibinfo{author}{\bibfnamefont{C.~W.} \bibnamefont{Xiao}},
  \bibinfo{author}{\bibfnamefont{J.}~\bibnamefont{Nieves}}, \bibnamefont{and}
  \bibinfo{author}{\bibfnamefont{E.}~\bibnamefont{Oset}}
  (\bibinfo{year}{2019}{\natexlab{a}}), \eprint{1904.01296}.

\bibitem[{\citenamefont{Shimizu et~al.}(2019)\citenamefont{Shimizu, Yamaguchi,
  and Harada}}]{Shimizu:2019ptd}
\bibinfo{author}{\bibfnamefont{Y.}~\bibnamefont{Shimizu}},
  \bibinfo{author}{\bibfnamefont{Y.}~\bibnamefont{Yamaguchi}},
  \bibnamefont{and} \bibinfo{author}{\bibfnamefont{M.}~\bibnamefont{Harada}}
  (\bibinfo{year}{2019}), \eprint{1904.00587}.

\bibitem[{\citenamefont{Guo and Oller}(2019)}]{Guo:2019kdc}
\bibinfo{author}{\bibfnamefont{Z.-H.} \bibnamefont{Guo}} \bibnamefont{and}
  \bibinfo{author}{\bibfnamefont{J.~A.} \bibnamefont{Oller}},
  \bibinfo{journal}{Phys. Lett.} \textbf{\bibinfo{volume}{B793}},
  \bibinfo{pages}{144} (\bibinfo{year}{2019}), \eprint{1904.00851}.

\bibitem[{\citenamefont{Wu et~al.}(2010)\citenamefont{Wu, Molina, Oset, and
  Zou}}]{Wu:2010jy}
\bibinfo{author}{\bibfnamefont{J.-J.} \bibnamefont{Wu}},
  \bibinfo{author}{\bibfnamefont{R.}~\bibnamefont{Molina}},
  \bibinfo{author}{\bibfnamefont{E.}~\bibnamefont{Oset}}, \bibnamefont{and}
  \bibinfo{author}{\bibfnamefont{B.~S.} \bibnamefont{Zou}},
  \bibinfo{journal}{Phys. Rev. Lett.} \textbf{\bibinfo{volume}{105}},
  \bibinfo{pages}{232001} (\bibinfo{year}{2010}), \eprint{1007.0573}.

\bibitem[{\citenamefont{Wu et~al.}(2011)\citenamefont{Wu, Molina, Oset, and
  Zou}}]{Wu:2010vk}
\bibinfo{author}{\bibfnamefont{J.-J.} \bibnamefont{Wu}},
  \bibinfo{author}{\bibfnamefont{R.}~\bibnamefont{Molina}},
  \bibinfo{author}{\bibfnamefont{E.}~\bibnamefont{Oset}}, \bibnamefont{and}
  \bibinfo{author}{\bibfnamefont{B.~S.} \bibnamefont{Zou}},
  \bibinfo{journal}{Phys. Rev.} \textbf{\bibinfo{volume}{C84}},
  \bibinfo{pages}{015202} (\bibinfo{year}{2011}), \eprint{1011.2399}.

\bibitem[{\citenamefont{Xiao et~al.}(2013)\citenamefont{Xiao, Nieves, and
  Oset}}]{Xiao:2013yca}
\bibinfo{author}{\bibfnamefont{C.~W.} \bibnamefont{Xiao}},
  \bibinfo{author}{\bibfnamefont{J.}~\bibnamefont{Nieves}}, \bibnamefont{and}
  \bibinfo{author}{\bibfnamefont{E.}~\bibnamefont{Oset}},
  \bibinfo{journal}{Phys. Rev.} \textbf{\bibinfo{volume}{D88}},
  \bibinfo{pages}{056012} (\bibinfo{year}{2013}), \eprint{1304.5368}.

\bibitem[{\citenamefont{Karliner and Rosner}(2015)}]{Karliner:2015ina}
\bibinfo{author}{\bibfnamefont{M.}~\bibnamefont{Karliner}} \bibnamefont{and}
  \bibinfo{author}{\bibfnamefont{J.~L.} \bibnamefont{Rosner}},
  \bibinfo{journal}{Phys. Rev. Lett.} \textbf{\bibinfo{volume}{115}},
  \bibinfo{pages}{122001} (\bibinfo{year}{2015}), \eprint{1506.06386}.

\bibitem[{\citenamefont{Wang et~al.}(2011)\citenamefont{Wang, Huang, Zhang, and
  Zou}}]{Wang:2011rga}
\bibinfo{author}{\bibfnamefont{W.~L.} \bibnamefont{Wang}},
  \bibinfo{author}{\bibfnamefont{F.}~\bibnamefont{Huang}},
  \bibinfo{author}{\bibfnamefont{Z.~Y.} \bibnamefont{Zhang}}, \bibnamefont{and}
  \bibinfo{author}{\bibfnamefont{B.~S.} \bibnamefont{Zou}},
  \bibinfo{journal}{Phys. Rev.} \textbf{\bibinfo{volume}{C84}},
  \bibinfo{pages}{015203} (\bibinfo{year}{2011}), \eprint{1101.0453}.

\bibitem[{\citenamefont{Yang et~al.}(2012)\citenamefont{Yang, Sun, He, Liu, and
  Zhu}}]{Yang:2011wz}
\bibinfo{author}{\bibfnamefont{Z.-C.} \bibnamefont{Yang}},
  \bibinfo{author}{\bibfnamefont{Z.-F.} \bibnamefont{Sun}},
  \bibinfo{author}{\bibfnamefont{J.}~\bibnamefont{He}},
  \bibinfo{author}{\bibfnamefont{X.}~\bibnamefont{Liu}}, \bibnamefont{and}
  \bibinfo{author}{\bibfnamefont{S.-L.} \bibnamefont{Zhu}},
  \bibinfo{journal}{Chin. Phys.} \textbf{\bibinfo{volume}{C36}},
  \bibinfo{pages}{6} (\bibinfo{year}{2012}), \eprint{1105.2901}.

\bibitem[{\citenamefont{Eides et~al.}(2019)\citenamefont{Eides, Petrov, and
  Polyakov}}]{Eides:2019tgv}
\bibinfo{author}{\bibfnamefont{M.~I.} \bibnamefont{Eides}},
  \bibinfo{author}{\bibfnamefont{V.~Y.} \bibnamefont{Petrov}},
  \bibnamefont{and} \bibinfo{author}{\bibfnamefont{M.~V.}
  \bibnamefont{Polyakov}} (\bibinfo{year}{2019}), \eprint{1904.11616}.

\bibitem[{\citenamefont{Ali and Parkhomenko}(2019)}]{Ali:2019npk}
\bibinfo{author}{\bibfnamefont{A.}~\bibnamefont{Ali}} \bibnamefont{and}
  \bibinfo{author}{\bibfnamefont{A.~{\relax Ya}.} \bibnamefont{Parkhomenko}},
  \bibinfo{journal}{Phys. Lett.} \textbf{\bibinfo{volume}{B793}},
  \bibinfo{pages}{365} (\bibinfo{year}{2019}), \eprint{1904.00446}.

\bibitem[{\citenamefont{Wang}(2019)}]{Wang:2019got}
\bibinfo{author}{\bibfnamefont{Z.-G.} \bibnamefont{Wang}}
  (\bibinfo{year}{2019}), \eprint{1905.02892}.

\bibitem[{\citenamefont{Cheng and Liu}(2019)}]{Cheng:2019obk}
\bibinfo{author}{\bibfnamefont{J.-B.} \bibnamefont{Cheng}} \bibnamefont{and}
  \bibinfo{author}{\bibfnamefont{Y.-R.} \bibnamefont{Liu}}
  (\bibinfo{year}{2019}), \eprint{1905.08605}.

\bibitem[{\citenamefont{Aaij et~al.}(2015)}]{Aaij:2015tga}
\bibinfo{author}{\bibfnamefont{R.}~\bibnamefont{Aaij}} \bibnamefont{et~al.}
  (\bibinfo{collaboration}{LHCb}), \bibinfo{journal}{Phys. Rev. Lett.}
  \textbf{\bibinfo{volume}{115}}, \bibinfo{pages}{072001}
  (\bibinfo{year}{2015}), \eprint{1507.03414}.

\bibitem[{\citenamefont{Yamaguchi et~al.}(2019)\citenamefont{Yamaguchi,
  Garcia-Tecocoatzi, Giachino, Hosaka, Santopinto, Takeuchi, and
  Takizawa}}]{Yamaguchi:2019seo}
\bibinfo{author}{\bibfnamefont{Y.}~\bibnamefont{Yamaguchi}},
  \bibinfo{author}{\bibfnamefont{H.}~\bibnamefont{Garcia-Tecocoatzi}},
  \bibinfo{author}{\bibfnamefont{A.}~\bibnamefont{Giachino}},
  \bibinfo{author}{\bibfnamefont{A.}~\bibnamefont{Hosaka}},
  \bibinfo{author}{\bibfnamefont{E.}~\bibnamefont{Santopinto}},
  \bibinfo{author}{\bibfnamefont{S.}~\bibnamefont{Takeuchi}}, \bibnamefont{and}
  \bibinfo{author}{\bibfnamefont{M.}~\bibnamefont{Takizawa}}
  (\bibinfo{year}{2019}), \eprint{1907.04684}.

\bibitem[{\citenamefont{Pavon~Valderrama}(2019{\natexlab{a}})}]{Valderrama:2019chc}
\bibinfo{author}{\bibfnamefont{M.}~\bibnamefont{Pavon~Valderrama}}
  (\bibinfo{year}{2019}{\natexlab{a}}), \eprint{1907.05294}.

\bibitem[{\citenamefont{Liu et~al.}(2019{\natexlab{b}})\citenamefont{Liu, Wu,
  Sánchez~Sánchez, Valderrama, Geng, and Xie}}]{Liu:2019zvb}
\bibinfo{author}{\bibfnamefont{M.-Z.} \bibnamefont{Liu}},
  \bibinfo{author}{\bibfnamefont{T.-W.} \bibnamefont{Wu}},
  \bibinfo{author}{\bibfnamefont{M.}~\bibnamefont{Sánchez~Sánchez}},
  \bibinfo{author}{\bibfnamefont{M.~P.} \bibnamefont{Valderrama}},
  \bibinfo{author}{\bibfnamefont{L.-S.} \bibnamefont{Geng}}, \bibnamefont{and}
  \bibinfo{author}{\bibfnamefont{J.-J.} \bibnamefont{Xie}}
  (\bibinfo{year}{2019}{\natexlab{b}}), \eprint{1907.06093}.

\bibitem[{\citenamefont{Sugiura et~al.}(2019)\citenamefont{Sugiura, Ikeda, and
  Ishii}}]{Sugiura:2019pye}
\bibinfo{author}{\bibfnamefont{T.}~\bibnamefont{Sugiura}},
  \bibinfo{author}{\bibfnamefont{Y.}~\bibnamefont{Ikeda}}, \bibnamefont{and}
  \bibinfo{author}{\bibfnamefont{N.}~\bibnamefont{Ishii}},
  \bibinfo{journal}{PoS} \textbf{\bibinfo{volume}{LATTICE2018}},
  \bibinfo{pages}{093} (\bibinfo{year}{2019}), \eprint{1905.02336}.

\bibitem[{\citenamefont{Skerbis and Prelovsek}(2019)}]{Skerbis:2018lew}
\bibinfo{author}{\bibfnamefont{U.}~\bibnamefont{Skerbis}} \bibnamefont{and}
  \bibinfo{author}{\bibfnamefont{S.}~\bibnamefont{Prelovsek}},
  \bibinfo{journal}{Phys. Rev.} \textbf{\bibinfo{volume}{D99}},
  \bibinfo{pages}{094505} (\bibinfo{year}{2019}), \eprint{1811.02285}.

\bibitem[{\citenamefont{Isgur and Wise}(1989)}]{Isgur:1989vq}
\bibinfo{author}{\bibfnamefont{N.}~\bibnamefont{Isgur}} \bibnamefont{and}
  \bibinfo{author}{\bibfnamefont{M.~B.} \bibnamefont{Wise}},
  \bibinfo{journal}{Phys. Lett.} \textbf{\bibinfo{volume}{B232}},
  \bibinfo{pages}{113} (\bibinfo{year}{1989}).

\bibitem[{\citenamefont{Isgur and Wise}(1990)}]{Isgur:1989ed}
\bibinfo{author}{\bibfnamefont{N.}~\bibnamefont{Isgur}} \bibnamefont{and}
  \bibinfo{author}{\bibfnamefont{M.~B.} \bibnamefont{Wise}},
  \bibinfo{journal}{Phys. Lett.} \textbf{\bibinfo{volume}{B237}},
  \bibinfo{pages}{527} (\bibinfo{year}{1990}).

\bibitem[{\citenamefont{Nieves and Valderrama}(2012)}]{Nieves:2012tt}
\bibinfo{author}{\bibfnamefont{J.}~\bibnamefont{Nieves}} \bibnamefont{and}
  \bibinfo{author}{\bibfnamefont{M.~P.} \bibnamefont{Valderrama}},
  \bibinfo{journal}{Phys. Rev.} \textbf{\bibinfo{volume}{D86}},
  \bibinfo{pages}{056004} (\bibinfo{year}{2012}), \eprint{1204.2790}.

\bibitem[{\citenamefont{Guo et~al.}(2013{\natexlab{a}})\citenamefont{Guo,
  Hidalgo-Duque, Nieves, and Valderrama}}]{Guo:2013xga}
\bibinfo{author}{\bibfnamefont{F.-K.} \bibnamefont{Guo}},
  \bibinfo{author}{\bibfnamefont{C.}~\bibnamefont{Hidalgo-Duque}},
  \bibinfo{author}{\bibfnamefont{J.}~\bibnamefont{Nieves}}, \bibnamefont{and}
  \bibinfo{author}{\bibfnamefont{M.~P.} \bibnamefont{Valderrama}},
  \bibinfo{journal}{Phys. Rev.} \textbf{\bibinfo{volume}{D88}},
  \bibinfo{pages}{054014} (\bibinfo{year}{2013}{\natexlab{a}}),
  \eprint{1305.4052}.

\bibitem[{\citenamefont{Xiao et~al.}(2019{\natexlab{b}})\citenamefont{Xiao,
  Nieves, and Oset}}]{Xiao:2019gjd}
\bibinfo{author}{\bibfnamefont{C.~W.} \bibnamefont{Xiao}},
  \bibinfo{author}{\bibfnamefont{J.}~\bibnamefont{Nieves}}, \bibnamefont{and}
  \bibinfo{author}{\bibfnamefont{E.}~\bibnamefont{Oset}}
  (\bibinfo{year}{2019}{\natexlab{b}}), \eprint{1906.09010}.

\bibitem[{\citenamefont{Peng et~al.}(2019)\citenamefont{Peng, Liu, Pan,
  Sánchez~Sánchez, and Valderrama}}]{Peng:2019wys}
\bibinfo{author}{\bibfnamefont{F.-Z.} \bibnamefont{Peng}},
  \bibinfo{author}{\bibfnamefont{M.-Z.} \bibnamefont{Liu}},
  \bibinfo{author}{\bibfnamefont{Y.-W.} \bibnamefont{Pan}},
  \bibinfo{author}{\bibfnamefont{M.}~\bibnamefont{Sánchez~Sánchez}},
  \bibnamefont{and} \bibinfo{author}{\bibfnamefont{M.~P.}
  \bibnamefont{Valderrama}} (\bibinfo{year}{2019}), \eprint{1907.05322}.

\bibitem[{\citenamefont{Bondar et~al.}(2011)\citenamefont{Bondar, Garmash,
  Milstein, Mizuk, and Voloshin}}]{Bondar:2011ev}
\bibinfo{author}{\bibfnamefont{A.~E.} \bibnamefont{Bondar}},
  \bibinfo{author}{\bibfnamefont{A.}~\bibnamefont{Garmash}},
  \bibinfo{author}{\bibfnamefont{A.~I.} \bibnamefont{Milstein}},
  \bibinfo{author}{\bibfnamefont{R.}~\bibnamefont{Mizuk}}, \bibnamefont{and}
  \bibinfo{author}{\bibfnamefont{M.~B.} \bibnamefont{Voloshin}},
  \bibinfo{journal}{Phys. Rev.} \textbf{\bibinfo{volume}{D84}},
  \bibinfo{pages}{054010} (\bibinfo{year}{2011}), \eprint{1105.4473}.

\bibitem[{\citenamefont{Mehen and Powell}(2011)}]{Mehen:2011yh}
\bibinfo{author}{\bibfnamefont{T.}~\bibnamefont{Mehen}} \bibnamefont{and}
  \bibinfo{author}{\bibfnamefont{J.~W.} \bibnamefont{Powell}},
  \bibinfo{journal}{Phys. Rev.} \textbf{\bibinfo{volume}{D84}},
  \bibinfo{pages}{114013} (\bibinfo{year}{2011}), \eprint{1109.3479}.

\bibitem[{\citenamefont{Guo et~al.}(2013{\natexlab{b}})\citenamefont{Guo,
  Hidalgo-Duque, Nieves, and Valderrama}}]{Guo:2013sya}
\bibinfo{author}{\bibfnamefont{F.-K.} \bibnamefont{Guo}},
  \bibinfo{author}{\bibfnamefont{C.}~\bibnamefont{Hidalgo-Duque}},
  \bibinfo{author}{\bibfnamefont{J.}~\bibnamefont{Nieves}}, \bibnamefont{and}
  \bibinfo{author}{\bibfnamefont{M.~P.} \bibnamefont{Valderrama}},
  \bibinfo{journal}{Phys. Rev.} \textbf{\bibinfo{volume}{D88}},
  \bibinfo{pages}{054007} (\bibinfo{year}{2013}{\natexlab{b}}),
  \eprint{1303.6608}.

\bibitem[{\citenamefont{Savage and Wise}(1990)}]{Savage:1990di}
\bibinfo{author}{\bibfnamefont{M.~J.} \bibnamefont{Savage}} \bibnamefont{and}
  \bibinfo{author}{\bibfnamefont{M.~B.} \bibnamefont{Wise}},
  \bibinfo{journal}{Phys. Lett.} \textbf{\bibinfo{volume}{B248}},
  \bibinfo{pages}{177} (\bibinfo{year}{1990}).

\bibitem[{\citenamefont{Sakai et~al.}(2019)\citenamefont{Sakai, Jing, and
  Guo}}]{Sakai:2019qph}
\bibinfo{author}{\bibfnamefont{S.}~\bibnamefont{Sakai}},
  \bibinfo{author}{\bibfnamefont{H.-J.} \bibnamefont{Jing}}, \bibnamefont{and}
  \bibinfo{author}{\bibfnamefont{F.-K.} \bibnamefont{Guo}}
  (\bibinfo{year}{2019}), \eprint{1907.03414}.

\bibitem[{\citenamefont{Liu et~al.}(2018{\natexlab{a}})\citenamefont{Liu, Peng,
  Sánchez~Sánchez, and Valderrama}}]{Liu:2018zzu}
\bibinfo{author}{\bibfnamefont{M.-Z.} \bibnamefont{Liu}},
  \bibinfo{author}{\bibfnamefont{F.-Z.} \bibnamefont{Peng}},
  \bibinfo{author}{\bibfnamefont{M.}~\bibnamefont{Sánchez~Sánchez}},
  \bibnamefont{and} \bibinfo{author}{\bibfnamefont{M.~P.}
  \bibnamefont{Valderrama}}, \bibinfo{journal}{Phys. Rev.}
  \textbf{\bibinfo{volume}{D98}}, \bibinfo{pages}{114030}
  (\bibinfo{year}{2018}{\natexlab{a}}), \eprint{1811.03992}.

\bibitem[{\citenamefont{Junnarkar and Mathur}(2019)}]{Junnarkar:2019equ}
\bibinfo{author}{\bibfnamefont{P.}~\bibnamefont{Junnarkar}} \bibnamefont{and}
  \bibinfo{author}{\bibfnamefont{N.}~\bibnamefont{Mathur}}
  (\bibinfo{year}{2019}), \eprint{1906.06054}.

\bibitem[{\citenamefont{Liu et~al.}(2018{\natexlab{b}})\citenamefont{Liu, Wu,
  Xie, Pavon~Valderrama, and Geng}}]{Liu:2018bkx}
\bibinfo{author}{\bibfnamefont{M.-Z.} \bibnamefont{Liu}},
  \bibinfo{author}{\bibfnamefont{T.-W.} \bibnamefont{Wu}},
  \bibinfo{author}{\bibfnamefont{J.-J.} \bibnamefont{Xie}},
  \bibinfo{author}{\bibfnamefont{M.}~\bibnamefont{Pavon~Valderrama}},
  \bibnamefont{and} \bibinfo{author}{\bibfnamefont{L.-S.} \bibnamefont{Geng}},
  \bibinfo{journal}{Phys. Rev.} \textbf{\bibinfo{volume}{D98}},
  \bibinfo{pages}{014014} (\bibinfo{year}{2018}{\natexlab{b}}),
  \eprint{1805.08384}.

\bibitem[{\citenamefont{Pavon~Valderrama}(2019{\natexlab{b}})}]{Valderrama:2019sid}
\bibinfo{author}{\bibfnamefont{M.}~\bibnamefont{Pavon~Valderrama}}
  (\bibinfo{year}{2019}{\natexlab{b}}), \eprint{1906.06491}.

\bibitem[{\citenamefont{Valderrama}(2012)}]{Valderrama:2012jv}
\bibinfo{author}{\bibfnamefont{M.~P.} \bibnamefont{Valderrama}},
  \bibinfo{journal}{Phys. Rev.} \textbf{\bibinfo{volume}{D85}},
  \bibinfo{pages}{114037} (\bibinfo{year}{2012}), \eprint{1204.2400}.

\bibitem[{\citenamefont{Lu et~al.}(2019)\citenamefont{Lu, Geng, and
  Valderrama}}]{Lu:2017dvm}
\bibinfo{author}{\bibfnamefont{J.-X.} \bibnamefont{Lu}},
  \bibinfo{author}{\bibfnamefont{L.-S.} \bibnamefont{Geng}}, \bibnamefont{and}
  \bibinfo{author}{\bibfnamefont{M.~P.} \bibnamefont{Valderrama}},
  \bibinfo{journal}{Phys. Rev.} \textbf{\bibinfo{volume}{D99}},
  \bibinfo{pages}{074026} (\bibinfo{year}{2019}), \eprint{1706.02588}.

\bibitem[{\citenamefont{Hu and Mehen}(2006)}]{Hu:2005gf}
\bibinfo{author}{\bibfnamefont{J.}~\bibnamefont{Hu}} \bibnamefont{and}
  \bibinfo{author}{\bibfnamefont{T.}~\bibnamefont{Mehen}},
  \bibinfo{journal}{Phys. Rev.} \textbf{\bibinfo{volume}{D73}},
  \bibinfo{pages}{054003} (\bibinfo{year}{2006}), \eprint{hep-ph/0511321}.

\bibitem[{\citenamefont{Cohen and Hohler}(2006)}]{Cohen:2006jg}
\bibinfo{author}{\bibfnamefont{T.~D.} \bibnamefont{Cohen}} \bibnamefont{and}
  \bibinfo{author}{\bibfnamefont{P.~M.} \bibnamefont{Hohler}},
  \bibinfo{journal}{Phys. Rev.} \textbf{\bibinfo{volume}{D74}},
  \bibinfo{pages}{094003} (\bibinfo{year}{2006}), \eprint{hep-ph/0606084}.

\bibitem[{\citenamefont{Padmanath et~al.}(2015)\citenamefont{Padmanath,
  Edwards, Mathur, and Peardon}}]{Padmanath:2015jea}
\bibinfo{author}{\bibfnamefont{M.}~\bibnamefont{Padmanath}},
  \bibinfo{author}{\bibfnamefont{R.~G.} \bibnamefont{Edwards}},
  \bibinfo{author}{\bibfnamefont{N.}~\bibnamefont{Mathur}}, \bibnamefont{and}
  \bibinfo{author}{\bibfnamefont{M.}~\bibnamefont{Peardon}},
  \bibinfo{journal}{Phys. Rev.} \textbf{\bibinfo{volume}{D91}},
  \bibinfo{pages}{094502} (\bibinfo{year}{2015}), \eprint{1502.01845}.

\bibitem[{\citenamefont{Chen and Chiu}(2017)}]{Chen:2017kxr}
\bibinfo{author}{\bibfnamefont{Y.-C.} \bibnamefont{Chen}} \bibnamefont{and}
  \bibinfo{author}{\bibfnamefont{T.-W.} \bibnamefont{Chiu}}
  (\bibinfo{collaboration}{TWQCD}), \bibinfo{journal}{Phys. Lett.}
  \textbf{\bibinfo{volume}{B767}}, \bibinfo{pages}{193} (\bibinfo{year}{2017}),
  \eprint{1701.02581}.

\bibitem[{\citenamefont{Alexandrou and Kallidonis}(2017)}]{Alexandrou:2017xwd}
\bibinfo{author}{\bibfnamefont{C.}~\bibnamefont{Alexandrou}} \bibnamefont{and}
  \bibinfo{author}{\bibfnamefont{C.}~\bibnamefont{Kallidonis}},
  \bibinfo{journal}{Phys. Rev.} \textbf{\bibinfo{volume}{D96}},
  \bibinfo{pages}{034511} (\bibinfo{year}{2017}), \eprint{1704.02647}.

\bibitem[{\citenamefont{Mathur and Padmanath}(2019)}]{Mathur:2018rwu}
\bibinfo{author}{\bibfnamefont{N.}~\bibnamefont{Mathur}} \bibnamefont{and}
  \bibinfo{author}{\bibfnamefont{M.}~\bibnamefont{Padmanath}},
  \bibinfo{journal}{Phys. Rev.} \textbf{\bibinfo{volume}{D99}},
  \bibinfo{pages}{031501} (\bibinfo{year}{2019}), \eprint{1807.00174}.

\bibitem[{\citenamefont{Mehen and Mohapatra}(2019)}]{Mehen:2019cxn}
\bibinfo{author}{\bibfnamefont{T.~C.} \bibnamefont{Mehen}} \bibnamefont{and}
  \bibinfo{author}{\bibfnamefont{A.}~\bibnamefont{Mohapatra}}
  (\bibinfo{year}{2019}), \eprint{1905.06965}.

\end{thebibliography}

\end{document}